\documentclass[10pt,twocolumn]{article}
\usepackage{naturestyle}

\begin{document}

\twocolumn[
\begin{@twocolumnfalse}
{\fontsize{20}{23}\selectfont\bfseries\raggedright
Interpretable physics-informed retrieval-augmented generation language model for end-to-end inorganic crystal synthesis planning
\par}
\vspace{8pt}

{\fontsize{10.5}{13}\selectfont
Wei-Jian Jiang\textsuperscript{1},
Ye-Nan Sha\textsuperscript{1},
Hui Guo\textsuperscript{1},
Jie Chen\textsuperscript{1},
Yu-Cai Liang\textsuperscript{1},
Ke Zhou\textsuperscript{1},
Qi-Long Gao\textsuperscript{2,*},
Dong-Lin Han\textsuperscript{1,*},
Xin-Gao Gong\textsuperscript{3,4,*},
Wan-Jian Yin\textsuperscript{1,4,*}
\par}
\vspace{6pt}

{\fontsize{8.7}{10.5}\selectfont
\textsuperscript{1}College of Energy, Soochow Institute for Energy and Materials Innovation (SIEMIS), and Jiangsu Provincial Key Laboratory for Advanced Carbon Materials and Wearable Energy Technologies, Soochow University, Suzhou 215006, China.\\
\textsuperscript{2}Key Laboratory of Materials Physics, Ministry of Education, School of Physics, Zhengzhou University, Zhengzhou 450001, China.\\
\textsuperscript{3}Key Laboratory of Computational Physical Sciences (Ministry of Education), Department of Physics, Fudan University, Shanghai 200433, China.\\
\textsuperscript{4}Hefei National Laboratory, Hefei 230088, China.\\
\textsuperscript{*}Correspondence:
\href{mailto:qilonggao@zzu.edu.cn}{qilonggao@zzu.edu.cn};
\href{mailto:dlhan@suda.edu.cn}{dlhan@suda.edu.cn};
\href{mailto:xggong@fudan.edu.cn}{xggong@fudan.edu.cn};
\href{mailto:wjyin@suda.edu.cn}{wjyin@suda.edu.cn}.
\par}
\vspace{9pt}

\noindent\textbf{Abstract}\quad
\textbf{Synthesis planning for inorganic materials, i.e., predicting both synthesizability and viable synthetic routes, remains a grand challenge in materials science. Its difficulty lies in coupling two perspectives that have long evolved in isolation: the microscopic, computationally accessible regime of thermodynamic convex-hull energetics, and the macroscopic, experimentally governed regime of synthesis methods, precursors, and processing conditions. Here, we develop an interpretable Physics-Informed Retrieval-Augmented Generation Language Model (PIRAG-LM) for end-to-end inorganic crystal synthesis planning. We first construct a material-centered Structured Synthesis Knowledge Base (SSKB) of 13,820 experimentally synthesized inorganic crystals, in which literature-derived synthesis information is organized around each material as route-level records. We then design a physics-based similarity module that retrieves historical precedents through three complementary descriptors, namely chemical, structural, and thermodynamic similarity, and couple it with a structured LLM reasoning module that proposes candidate routes, specifies precursors and processing conditions, and scores each route across thermodynamic feasibility, kinetics, and accessibility. PIRAG-LM achieves 91.4\% accuracy in synthesis-method prediction, compared with 72.1\% for the LLM alone, and generalizes robustly to materials reported after the knowledge-based cutoff. Because PIRAG-LM is built on retrieval rather than parametric memorization, its accuracy can be continually improved simply by enriching the SSKB, which takes effect immediately without any retraining or fine-tuning of the language model, making the framework convenient and fast to extend. Guided by PIRAG-LM, we experimentally synthesize five new compounds BaMo\textsubscript{0.3}In\textsubscript{0.7}O\textsubscript{2.95}, BaNb\textsubscript{0.4}In\textsubscript{0.6}O\textsubscript{2.9}, Hg[B(CN)\textsubscript{4}]\textsubscript{2}, CoCo(CN)\textsubscript{6}, and SrNb\textsubscript{2}Fe\textsubscript{2}(PO\textsubscript{4})\textsubscript{6} via both solid-state and solution routes, demonstrating its value as an interpretable machine-learning approach that narrows the gap between computational materials discovery and experimental realization.}
\vspace{12pt}
\end{@twocolumnfalse}
]

\section{Introduction}

The integration of computational materials science and artificial intelligence has greatly expanded the space of candidate inorganic materials. High-throughput screening and generative models have populated databases such as Materials Project, OQMD, and GNoMe with materials on the order of 10\textsuperscript{6} \cite{ref1,ref2,ref3,ref4}. However, only a small fraction has been experimentally synthesized, leaving a major gap between computational prediction and experimental realization.

Physics-driven approaches have sought to address this gap using synthesizability metrics. Convex-hull energy and the amorphous limit provide interpretable measures of thermodynamic stability\cite{ref5,ref6,ref7}, while kinetic descriptors such as the basin of attraction and transition barrier offer complementary insights from the potential-energy landscape\cite{ref8}. Nevertheless, thermodynamic stability does not guarantee synthesis, and metastable phases can also be experimentally accessible. More importantly, these metrics do not directly provide actionable synthesis information, such as suitable precursors, temperatures, atmospheres, or processing routes.

Because such information is primarily reported in the literature as natural language, LLMs have recently been applied to synthesis planning. CSLLM and StructGPT-FT demonstrated LLM-based synthesizability classification and post hoc interpretation\cite{ref9,ref10}, whereas literature-grounded retrieval systems have shown promise for precursor recommendation and scientific reasoning\cite{ref11,ref12,ref13}. These studies indicate that LLMs can connect published synthesis knowledge to new materials, but most existing methods either frame synthesizability as a material-level classification problem or retrieve literature primarily through textual similarity.

Thus, current approaches face a complementary limitation: physics-based methods are interpretable but lack route-level experimental guidance\cite{ref14}, whereas LLM-based methods can propose routes but often provide opaque reasoning. Directly supplying full-text literature to an LLM further introduces redundant context and can obscure the association among precursors, processing conditions, and target phases. A synthesis-planning framework should therefore organize literature into structured, route-level records and retrieve precedents using physically meaningful material similarities, enabling both actionable and interpretable synthesis recommendations.

In this work, we reformulate inorganic material synthesizability from a binary classification problem, determined solely by the material itself, into a route-dependent scoring problem defined over multiple possible synthesis pathways. Building on this reformulation, we develop PIRAG-LM, a physics-informed retrieval-augmented generation language model for end-to-end inorganic crystal synthesis planning. The framework integrates three components: (i) a material-centered Structured Synthesis Knowledge Base (SSKB) containing 13,820 experimentally synthesized inorganic crystals curated from MP and ICSD\cite{ref1,ref2}, in which literature-derived synthesis information is organized around each material as route-level records; (ii) a physics-based similarity module that retrieves precedents through three complementary descriptors, \emph{i.e.}, chemical (CS), structural (SS), and thermodynamic (TS) similarity, ensuring retrieval grounded in physical analogies rather than textual resemblance; and (iii) a structured LLM reasoning module that proposes candidate routes, specifies precursors and processing conditions, and scores each route across thermodynamic feasibility, kinetics, and accessibility. Because every recommendation is explicitly linked to retrieved precedents and physical descriptors, the reasoning process is inherently traceable and interpretable.

We validate PIRAG-LM on representative inorganic systems curated from the MP database, where it accurately predicts synthesis methods, precursors, and key processing conditions, while flagging potential experimental challenges. A time-axis evaluation using a pre-2021 knowledge base to predict materials reported during 2021--2026 confirms strong generalization beyond the training corpus. Ablation studies show that the full CS+SS+TS retrieval achieves 91.4\% accuracy, compared with 72.1\% for the LLM alone, verifying that the performance gain stems from physics-informed retrieval. Guided by PIRAG-LM, we further experimentally synthesize five new compounds BaMo\textsubscript{0.3}In\textsubscript{0.7}O\textsubscript{2.95}, BaNb\textsubscript{0.4}In\textsubscript{0.6}O\textsubscript{2.9}, Hg{[}B(CN)\textsubscript{4}{]}\textsubscript{2}, CoCo(CN)\textsubscript{6}, and SrNb\textsubscript{2}Fe\textsubscript{2}(PO\textsubscript{4})\textsubscript{6}, demonstrating its practical value as an interpretable model. Overall, this work shows that the gap between computational discovery and experimental realization can be narrowed by combining physical descriptors, historical synthesis knowledge, and language-model reasoning, shifting synthesizability prediction from binary classification toward interpretable, evidence-grounded synthesis planning.

\section{Results}

\subsection{Structured synthesis knowledge base (SSKB)}

\begin{figure*}[!t]
  \centering
  \includegraphics[width=\textwidth,height=0.76\textheight,keepaspectratio]{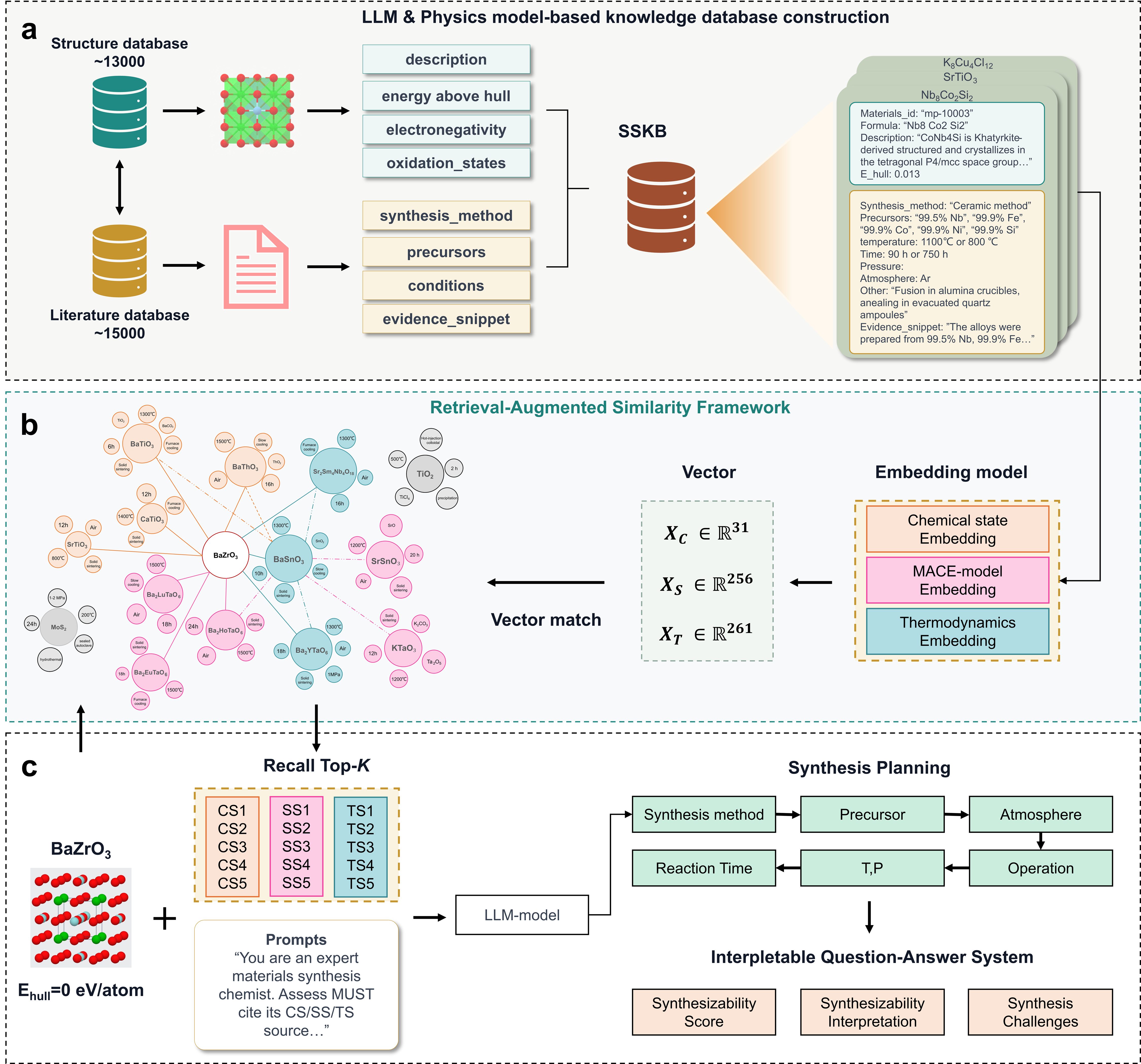}
  \caption{PIRAG-LM framework for inorganic-crystal synthesis planning. (a) Construction of the material-centered Structured Synthesis Knowledge Base (SSKB) from crystal structures, literature-derived synthesis records, and chemical and thermodynamic descriptors. (b) Retrieval-Augmented Similarity Framework, in which CS, SS, and TS embeddings are matched against the SSKB. (c) Top-k physical precedents and their route-level records are supplied to the language model to generate synthesis methods, precursors, conditions, scores, interpretations, and anticipated challenges.}
  \label{fig:1}
\end{figure*}

To obtain material-centered and route-resolved structured synthesis knowledge, we mined synthesis-related information for 13,820 inorganic materials in the ICSD\cite{ref2}. We used an LLM (DeepSeek-V4) for this task. We first screened approximately 15,000 articles, removed theoretical and characterization-only papers, and retained experimental studies. From these papers, we extracted hundreds of thousands of synthesis-related sentences and used the LLM's language-understanding capabilities to classify them into categories including synthesis methods, synthesis conditions, precursors and other key synthesis information. Overall, our approach achieved more accurate information extraction (IE) than rule-based text-matching methods\cite{ref15} (Fig.~\ref{fig:1}a).

In addition to synthesis information, the database also needs to include structural, chemical, and physical information that facilitates LLM comprehension. For structural information, we adopted a natural language representation format, converting a CIF structure into natural language using Robocrystallographer\cite{ref16}. Compared with raw structure files such as CIF or POSCAR, this language-based representation is more suitable for LLM processing while retaining key crystallographic and local-structure information, including space group, structural prototype, coordination environment, and atomic arrangement\cite{ref17}. This approach has been widely used and proven to accurately reflect structural correlations in previous language model predictions\cite{ref10}. Chemical information is primarily used to assess elemental compatibility; the SSKB includes information that can be readily obtained from the structure, such as electronegativity, valence state, oxidation state, coordination, etc. Thermodynamic information is also crucial, because materials occupying different regions of the free-energy landscape often require different synthesis strategies. Metastable phases are frequently accessed through non-equilibrium or constraint-assisted routes, such as vapor-phase deposition, thin-film epitaxy, high-pressure synthesis, or rapid quenching, where kinetic trapping, substrate stabilization, or external thermodynamic handles can stabilize phases that are not the ambient ground state\cite{ref6,ref18}. In contrast, long-duration high-temperature solid-state reactions generally allow more extensive diffusion and phase equilibration, and therefore tend to favor thermodynamically stable or low-energy products, although kinetic and precursor-dependent effects can still lead to metastable phases\cite{ref19}. We directly obtained this thermodynamic information, as reflected by the energy above hull, from the MP database (Fig.~\ref{fig:1}a). In its final form, the SSKB resembles a material-centered knowledge graph, where each material is the core entity and synthesis methods, precursors, conditions, processing steps, constraints, and literature sources are linked as associated entities. This structure preserves synthesis information as route-level evidence connected to specific materials, enabling traceable retrieval and reasoning for synthesis planning.

To ensure accuracy and generalizability, it is essential to evaluate the coverage of the SSKB. Since our model is grounded in historical synthesis knowledge, it can only be reliable when the database spans a sufficiently broad region of the experimentally explored materials space. We therefore examined the coverage from three complementary perspectives: elemental diversity, structural distribution, and synthesis-method diversity. As shown in Fig.~\ref{fig:2}a, the database covers most chemically relevant elements in inorganic solids, indicating broad chemical diversity beyond narrow material families.

We further evaluated the structural coverage using the MP database as a benchmark. The MACE atomic foundation model was used to embed crystal structures into high-dimensional feature vectors, followed by UMAP dimensionality reduction for visualization\cite{ref20}. The results in Fig.~\ref{fig:2}b show that known synthesized structures broadly span the MP materials space, and the 13,820 materials included in our database also exhibit substantial coverage of this space. We summarized the synthesis methods extracted from the literature, as shown in Fig.~\ref{fig:2}c. The database contains the vast majority of commonly used inorganic synthesis routes, with solid-state sintering being the most frequent method. Together, these analyses demonstrate that our physics--knowledge database provides broad chemical, structural, and methodological coverage for retrieval-based synthesis planning.

\subsection{Retrieval-augmented similarity framework (RASF)}

PIRAG-LM formulates synthesis-route prediction as a precedent-guided reasoning problem. For a target material, the framework retrieves experimentally synthesized materials with relevant physical properties and uses their route-level records as evidence for LLM reasoning (Fig.~\ref{fig:1}a,b). Unlike conventional text-based RAG, which retrieves semantically similar passages, PIRAG-LM retrieves well-defined material objects through chemical, structural, and thermodynamic similarity. This physically grounded retrieval constrains the LLM with explicit precedents and makes each recommendation traceable to specific material analogues.

The three similarity spaces provide complementary route-relevant information. CS captures similarities in elemental chemistry and coordination preferences, SS captures shared structural motifs, and TS distinguishes materials with different energetic stability despite similar chemistry or structure. Thus, chemically related oxides or structurally related perovskites may share transferable synthesis strategies, whereas metastable phases, such as nitride perovskites, may require non-equilibrium or substrate-stabilized thin-film routes\cite{ref6,ref18,ref21}.

\begin{figure*}[!t]
  \centering
  \includegraphics[width=\textwidth,height=0.76\textheight,keepaspectratio]{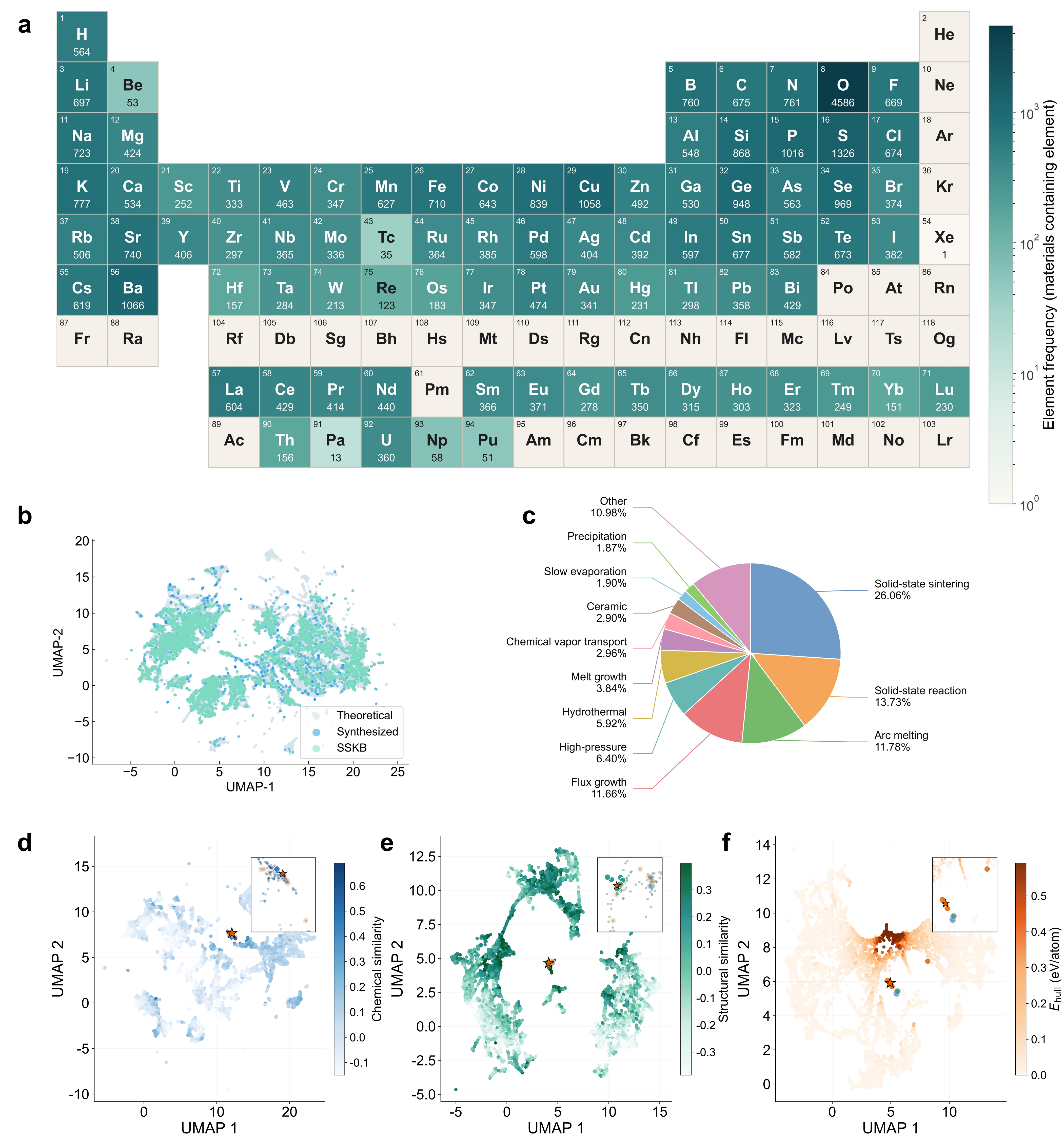}
  \caption{Coverage of the SSKB and physics-informed retrieval spaces. (a) Elemental occurrence across the SSKB. (b) UMAP projection comparing theoretical and experimentally synthesized structures from the Materials Project with structures included in the SSKB. (c) Distribution of synthesis-method families in the SSKB. (d--f) UMAP representations of the chemical, structural, and thermodynamic-state embedding spaces, respectively; the insets show the local neighborhoods retrieved for the example target.}
  \label{fig:2}
\end{figure*}

This material-level retrieval is enabled by the SSKB, which organizes literature into compact, route-resolved records linking composition, crystal structure, thermodynamic descriptors, precursors, conditions, synthesis routes, and source evidence. Chemical features, MACE-based structural embeddings\cite{ref22}, and a thermodynamic representation combining energy above hull with structural information are independently encoded for similarity search (Fig.~\ref{fig:2}d--f). Incorporating structural information into the thermodynamic representation avoids retrieving structurally unrelated materials with similar metastability. The resulting records provide focused and traceable evidence while reducing redundant context for the LLM; detailed encoder construction and retrieval process are provided in the \textbf{Methods}.

For a target material specified by a CIF file, PIRAG-LM computes its chemical and thermodynamic properties and generates embeddings in the chemical, structural, and thermodynamic spaces. The five nearest synthesized precedents from each space are retrieved, and their route-level synthesis records, structural descriptions, and physicochemical properties are supplied to the LLM as CS, SS, and TS evidence (Fig.~\ref{fig:1}b).

A structured prompt then guides the LLM to generate candidate synthesis routes, precursor sets, and processing conditions, and to evaluate each route in terms of thermodynamic feasibility, experimental accessibility, reaction kinetics, and amorphization tendency (Fig.~\ref{fig:1}c). Details of the prompt engineering are provided in the Supplementary Information. The resulting outputs include route-level difficulty scores, supporting rationales, and anticipated synthesis challenges.

\subsection{Prediction of synthesis routes, precursors and conditions}

\begin{figure*}[!t]
  \centering
  \includegraphics[width=\textwidth,height=0.76\textheight,keepaspectratio]{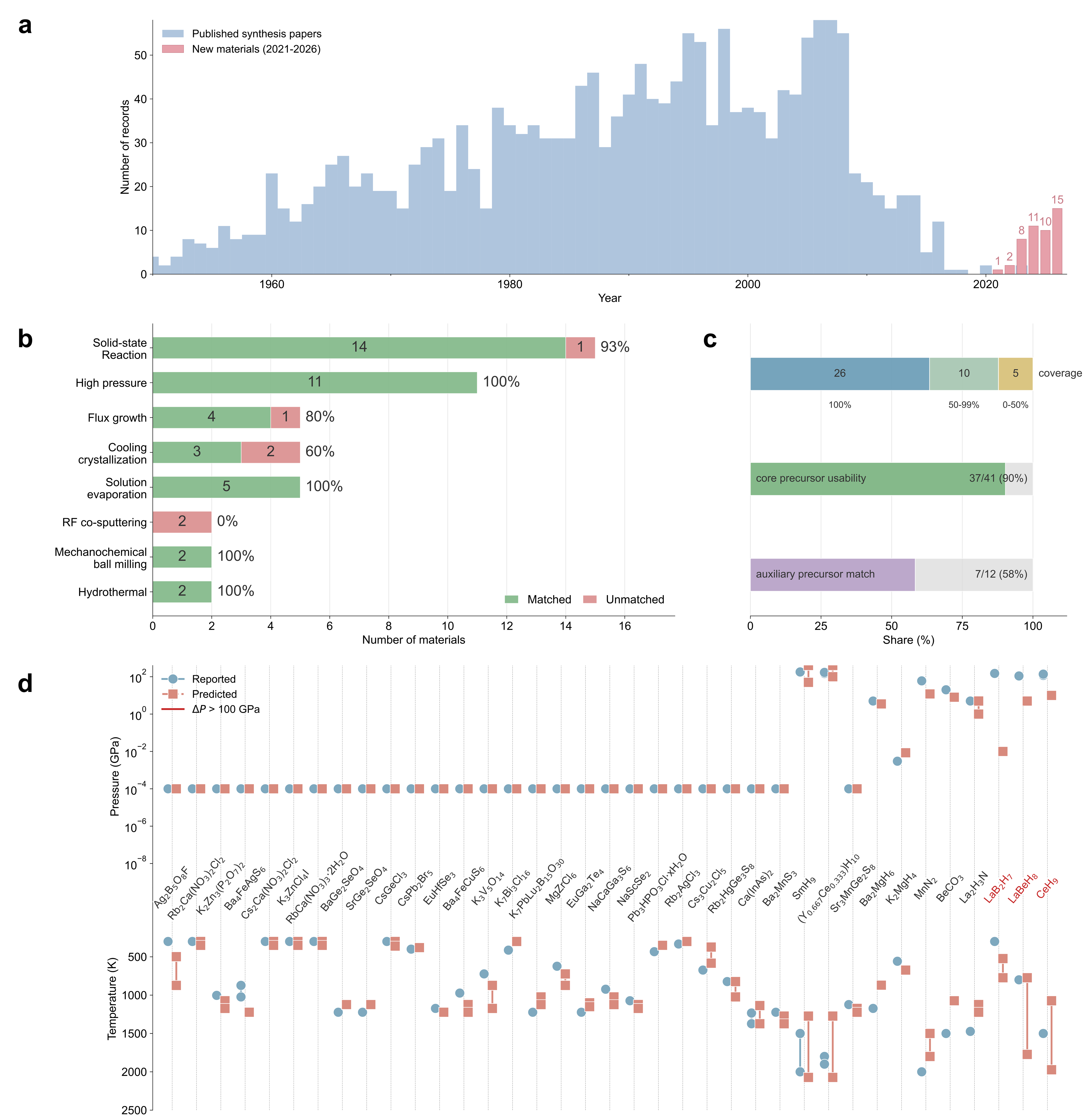}
  \caption{Time-axis evaluation of PIRAG-LM on recently reported materials. (a) Publication-year distribution of historical synthesis records and the 2021--2026 evaluation set. (b) Route-family matches and mismatches by synthesis method. (c) Precursor-coverage distribution, core-precursor usability, and auxiliary-precursor matching. (d) Reported and predicted synthesis temperatures and pressures; highlighted labels mark large temperature or pressure discrepancies.}
  \label{fig:3}
\end{figure*}

Because synthesis planning is a hierarchical task, we evaluated the framework in the same order in which an experimental synthesis plan is constructed: synthesis method, precursor set, and reaction conditions. The synthesis information in our database is derived primarily from materials and literature published before 2021; therefore, we adopted a time-axis-based evaluation strategy using the pre-2021 database to predict newly synthesized materials reported between 2021 and 2026 (Fig.~\ref{fig:3}a). In this evaluation, synthesis method prediction was assessed for all 47 recently reported materials. Precursor prediction was then evaluated for the subset with correctly predicted synthesis methods, because precursor choice is strongly method-dependent. Finally, temperature and pressure predictions were evaluated only for the 37 materials for which both the synthesis method and core precursor predictions were considered correct. Detailed criteria for determining the agreement between predicted and experimentally reported synthesis methods, precursor sets, temperatures, and pressures are described in the \textbf{Methods}.

We first evaluated synthesis method prediction. 41 of the 47 synthesis methods were correctly predicted, corresponding to an accuracy of 87.2\%. High-pressure synthesis (11/11), solution evaporation (5/5), hydrothermal synthesis (2/2), and mechanochemical ball milling (2/2) were predicted correctly in all evaluated cases. Solid-state reaction and flux growth achieved accuracies of 14/15 (93.3\%) and 4/5 (80.0\%), respectively, whereas cooling crystallization achieved 3/5 (60.0\%). Neither of the two RF co-sputtering cases was recovered (Fig.~\ref{fig:3}b).

The six method-level errors mainly involved specialized routes. Thin-film targets LaWN\textsubscript{3}\cite{ref23} and MgMoN\textsubscript{2}\cite{ref21} were misclassified as bulk flux-growth or solid-state routes, while Mg\textsubscript{4}Pt\textsubscript{3}H\textsubscript{6} was not recognized as requiring diamond-anvil-cell synthesis. The remaining errors involved flux-growth or cooling-crystallization systems. Overall, PIRAG-LM performs well for common bulk routes but is less reliable when synthesis depends on thin-film geometry, crystallization control, or ultrahigh-pressure stabilization.

For the 41 materials with correctly predicted synthesis methods, we assessed precursor prediction using three complementary criteria rather than exact string matching (Fig.~\ref{fig:3}c; see Methods). This distinction is necessary because a target phase can often be synthesized from chemically equivalent salts, oxides, hydrides, elements, or binary compounds. Requiring the predicted precursor list to be identical to one reported recipe would therefore classify viable chemical alternatives as errors. At the same time, merely supplying the correct elements is not sufficient if the proposed reagents are incompatible with the predicted route or omit a chemically necessary source or reactive building unit.

Precursor performance was evaluated using precursor coverage and core-precursor usability (\textbf{Methods}). Among the 41 method-correct cases, 26 achieved complete precursor coverage, 10 achieved 50--99\% coverage, and 5 achieved below 50\% coverage; thus, 36/41 cases (87.8\%) reached at least 50\% coverage. Core precursor sets were considered usable for 37/41 cases (90.2\%). The four rejected cases involved a deficient sulfur source for EuZnGeS\textsubscript{4}\cite{ref24}, failure to recover the Li\textsubscript{6}WN\textsubscript{4} + ZnBr\textsubscript{2} metathesis chemistry for Zn\textsubscript{3}WN\textsubscript{4}\cite{ref25}, replacement of the elemental Cs/V/Te flux with Cs\textsubscript{2}Te\textsubscript{3} + V for Cs\textsubscript{3}V\textsubscript{9}Te\textsubscript{13}\cite{ref26}, and a different nitrogen source and high-pressure mechanism for Sb\textsubscript{3}N\textsubscript{5}\cite{ref27}. These predictions were therefore counted as precursor errors rather than condition-prediction errors. Auxiliary-precursor matching was evaluated only for the 12 method-correct cases with experimentally identifiable route-specific additives, such as solvents, mineralizers, fluxes, ligands, or reactive media (\textbf{Methods}). Seven cases were matched, corresponding to 58.3\%. This lower rate reflects the greater dependence of auxiliary reagents on route-specific experimental implementation.

Reaction-condition prediction was assessed for the remaining 37 materials using the interval-distance metric described in the \textbf{Methods}, for which overlapping predicted and reported intervals give zero error. The temperature MAE was 91.8 K. The largest deviations were observed for BeCO\textsubscript{3}(427 K)\cite{ref28} and Ba\textsubscript{2}MgH\textsubscript{6}(303 K)\cite{ref29}, despite otherwise accepted method and precursor predictions, indicating that temperature transfer remains sensitive to precursor activation and high-pressure heating protocols.

The pressure MAE was 11.47 GPa and was dominated by three ultrahigh-pressure hydrides: LaB\textsubscript{2}H\textsubscript{7}\cite{ref30} (predicted 0.01 GPa versus reported 150 GPa), LaBeH\textsubscript{8}\cite{ref31} (5 versus 110 GPa), and CeH\textsubscript{9}\cite{ref32} (10 versus 118--137 GPa). These errors reflect failure to identify the required pressure regime rather than ordinary variation in experimental conditions.

Excluding these three pressure-regime failures reduced the pressure MAE to 1.81 GPa for the remaining 34 materials, although MnN\textsubscript{2}\cite{ref33} still showed a 48 GPa deviation. Thus, PIRAG-LM is reliable for most common routes and conventional precursor systems, while its main limitations involve specialized thin-film or crystallization routes and ultrahigh-pressure stabilization.

To further assess the robustness of PIRAG-LM beyond the temporally held-out cohort, we evaluated the framework on an additional strictly separated test set of 1,000 materials spanning all synthesis-reporting years. None of the target materials or their synthesis records were included in the SSKB. PIRAG-LM correctly predicted the synthesis method for 940 of 1,000 materials (94.0\%), usable precursor sets for 913 of the 940 method-correct cases (97.1\%), and synthesis conditions for 879 of the 913 precursor-correct cases (96.2\%). These results show that the framework maintains strong performance on a large-scale material-level held-out benchmark, while the preceding time-axis evaluation provides the more stringent test of temporal generalization.

Ablation experiments confirmed that the performance gain arises from physics-informed retrieval rather than the LLM alone (Table~\ref{tab:performance}). The LLM-only baseline achieved 72.1\% accuracy, whereas single-space retrieval provided limited improvement (62.8--79.1\%) and pairwise combinations reached 83.7--86.0\%. The full CS+SS+TS framework achieved the highest accuracy of 87.2\%, demonstrating that chemical, structural, and thermodynamic precedents provide complementary information for synthesis-route prediction.

\begin{table*}[t]
  \caption{Synthesis method prediction accuracy of the LLM without RAG and RAG systems with different configurations.}
  \label{tab:performance}
  \centering
  \begin{tabular}{@{}lr@{}}
    \toprule
    \textbf{Framework} & \textbf{Synthesis method accuracy} \\
    \midrule
    LLM-only & 72.1\% \\
    CS-LLM & 62.8\% \\
    SS-LLM & 79.1\% \\
    TS-LLM & 76.7\% \\
    CS+SS-LLM & 83.7\% \\
    CS+TS-LLM & 86.0\% \\
    SS+TS-LLM & 86.0\% \\
    CS+SS+TS-LLM & 87.2\% \\
    CS+SS+TS-LLM Enhanced & 91.4\% \\
    \bottomrule
  \end{tabular}
\end{table*}

\subsection{Performance enhanced by physics information and knowledge}

The preceding analyses demonstrate that PIRAG-LM performs reliably across the majority of synthesis-condition predictions, yet they also expose a small number of systematic errors that cannot be resolved by retrieval from ambient-pressure bulk precedents alone. Fundamentally, the predictive accuracy of PIRAG-LM is governed by the quality and completeness of its SSKB, whose information originates from two complementary sources: physics-based descriptors derived from first-principles calculations, and synthesis knowledge extracted from the experimental literature (Fig.~\ref{fig:1}a). Accordingly, the framework can be strengthened along these same two axes, by enriching its physical descriptors and by expanding its literature-derived synthesis knowledge. Because every prediction is explicitly linked to these sources, such errors can be diagnosed and corrected through targeted augmentation of the model input, without any retraining. Here, we demonstrate this extensibility through two representative failure modes, each addressed by augmenting one of the two information sources.

\begin{figure*}[!t]
  \centering
  \includegraphics[width=\textwidth,height=0.76\textheight,keepaspectratio]{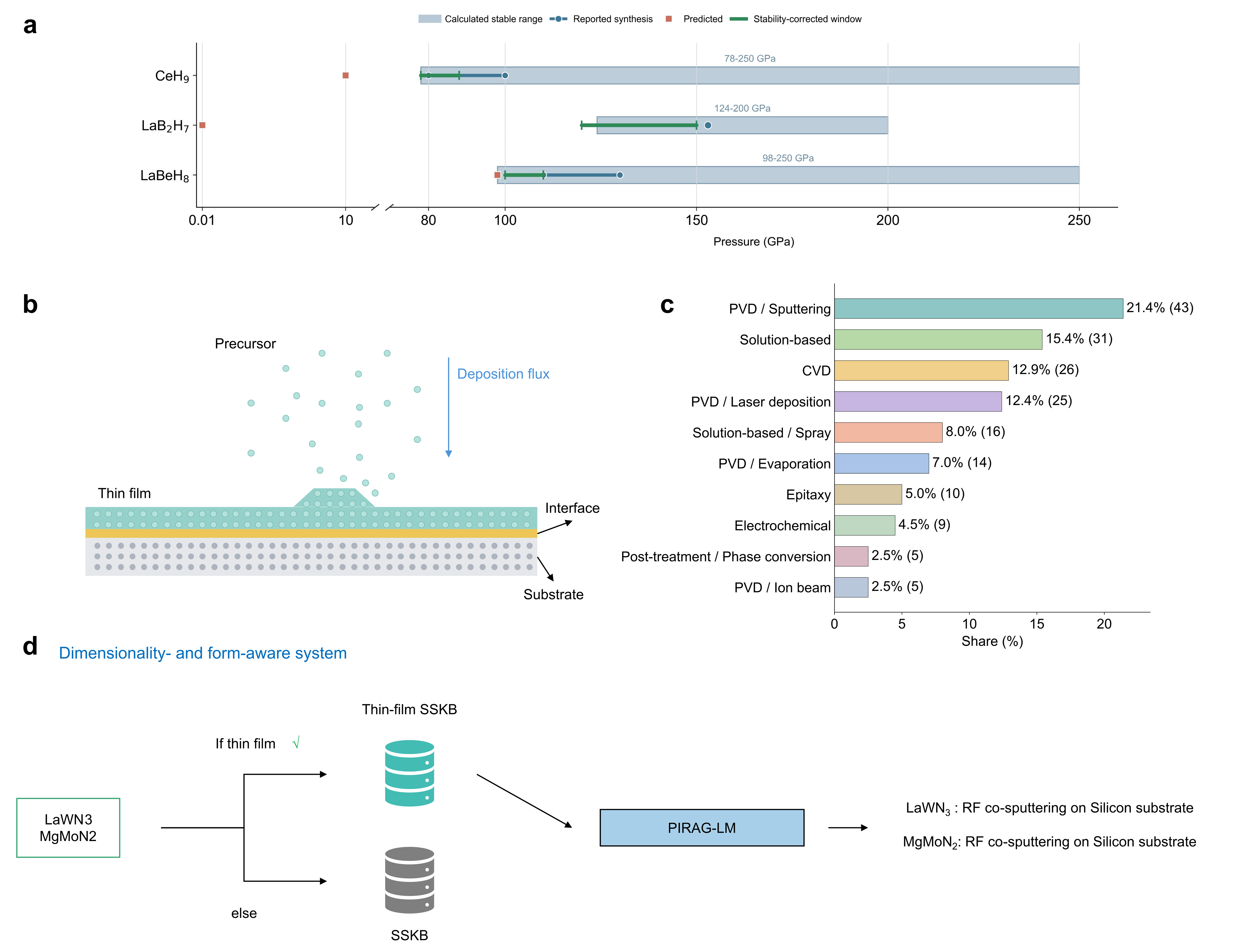}
  \caption{Physics- and form-aware correction of representative synthesis-prediction failures. (a) Calculated stability ranges, reported synthesis pressures, original predictions, and stability-corrected pressure windows for CeH\textsubscript{9}, LaB\textsubscript{2}H\textsubscript{7}, and LaBeH\textsubscript{8}. (b) Schematic of thin-film growth and substrate stabilization. (c) Distribution of the ten most frequent synthesis-method categories in the thin-film knowledge base; less frequent categories are omitted for clarity. (d) Dimensionality- and form-aware routing that directs thin-film targets to a dedicated SSKB and recovers RF co-sputtering routes for LaWN\textsubscript{3} and MgMoN\textsubscript{2}.}
  \label{fig:4}
\end{figure*}

The first failure mode was severe pressure underestimation for DAC-level hydrides. Because the SSKB contained only ambient-pressure convex-hull energies, it did not capture the ultrahigh-pressure stability of these phases and consequently biased predictions toward low pressures. Incorporating finite-pressure stability windows for CeH\textsubscript{9}\cite{ref32}, LaB\textsubscript{2}H\textsubscript{7}\cite{ref30}, and LaBeH\textsubscript{8}\cite{ref31} (78--250, 124--200, and 98--250 GPa, respectively) shifted the predicted pressure ranges into the experimentally relevant regime (Fig.~\ref{fig:4}a). Thus, pressure-prediction errors can be diagnosed and corrected by adding pressure-dependent thermodynamic information without retraining the language model.

The second failure mode was insufficient recognition of dimensionality-sensitive thin-film growth. Because the SSKB was dominated by bulk-synthesis records, retrieval favored conventional bulk routes even for thin-film targets. In thin-film growth, however, substrate selection, epitaxial stabilization, deposition flux, and non-equilibrium conditions can be essential for phase formation (Fig.~\ref{fig:4}b). We therefore added 201 thin-film synthesis records covering major deposition paradigms to a dedicated knowledge base; the ten most frequent synthesis-method categories are summarized in Fig.~\ref{fig:4}c, with less frequent categories omitted for clarity. When a target was identified as a thin film, PIRAG-LM preferentially retrieved these form-specific precedents and correctly assigned RF co-sputtering on Si substrates for LaWN\textsubscript{3}\cite{ref23} and MgMoN\textsubscript{2}\cite{ref21} (Fig.~\ref{fig:4}d), demonstrating that such errors can be corrected by expanding the SSKB without retraining. The enhanced framework achieved 91.4\% accuracy (Table~\ref{tab:performance}).

Together, these examples demonstrate the modular extensibility of PIRAG-LM: targeted additions to physical descriptors or synthesis records can correct systematic errors without retraining, highlighting its potential for synthesis planning in more complex material systems. This retrieval-based design allows PIRAG-LM to evolve continuously alongside the growth of physical and experimental knowledge, offering a scalable foundation for AI-assisted synthesis planning.

\subsection{Experimental validation of PIRAG-LM for synthesis planning}

\begin{figure*}[!t]
  \centering
  \includegraphics[width=\textwidth,height=0.76\textheight,keepaspectratio]{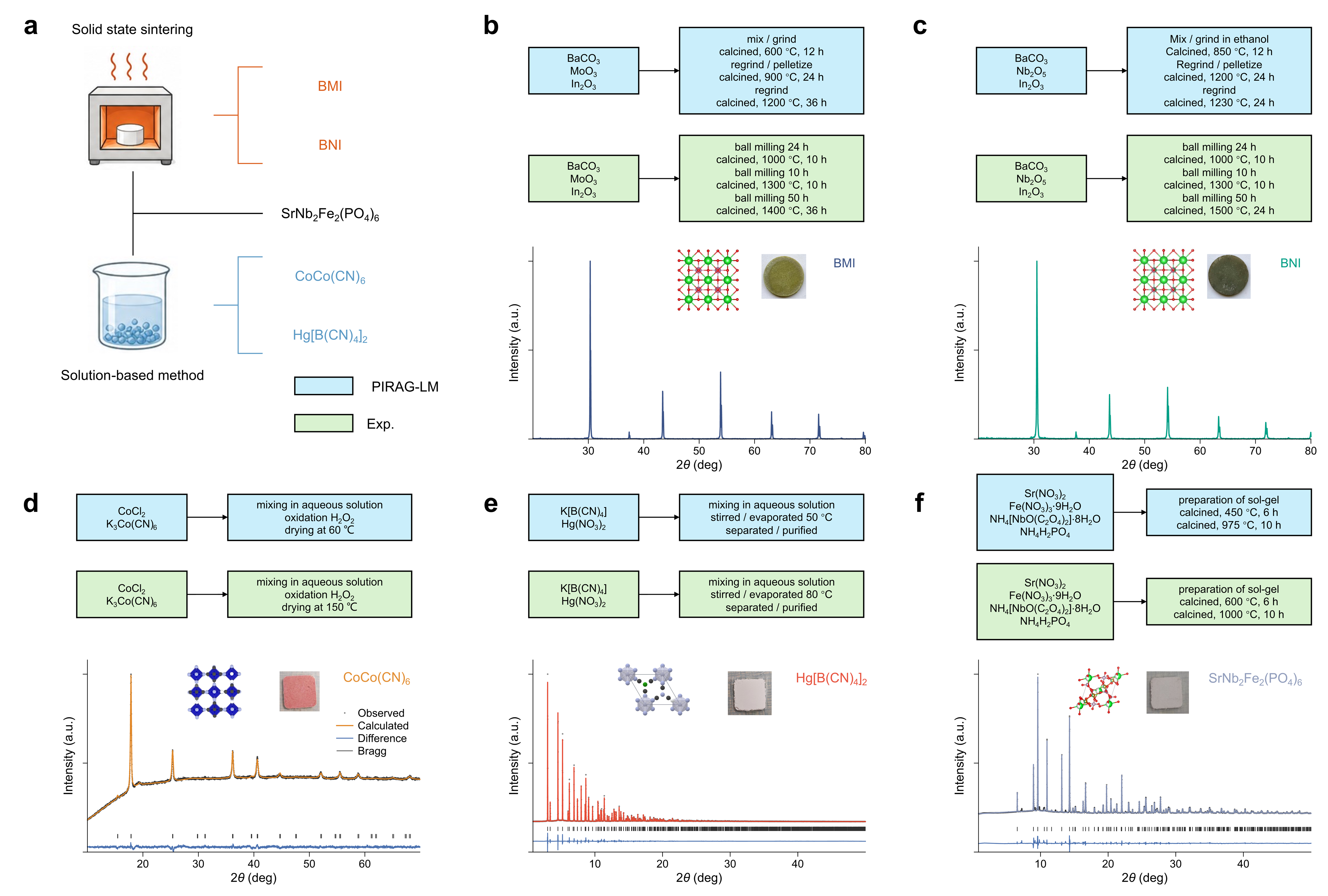}
  \caption{Experimental validation of PIRAG-LM synthesis plans. (a) Overview of the direct solid-state, sol--solid, and solution-phase synthesis target systems. (b, c) Predicted and experimentally implemented solid-state protocols for BaMo\textsubscript{0.3}In\textsubscript{0.7}O\textsubscript{2.95} and BaNb\textsubscript{0.4}In\textsubscript{0.6}O\textsubscript{2.9}, together with experimental powder X-ray diffraction (PXRD) patterns and target structures. (d, e) Corresponding protocol comparisons for CoCo(CN)\textsubscript{6} and Hg{[}B(CN)\textsubscript{4}{]}\textsubscript{2}, both prepared by solution-phase synthesis. (f) Protocol comparison and powder X-ray diffraction results for SrNb\textsubscript{2}Fe\textsubscript{2}(PO\textsubscript{4})\textsubscript{6}, synthesized by the sol--solid route. Blue boxes denote PIRAG-LM recommendations and green boxes denote experimentally implemented procedures.}
  \label{fig:5}
\end{figure*}

To evaluate the ability of PIRAG-LM to move from theoretical prediction to experimental realization, we went beyond the retrospective benchmarking described above and performed prospective experimental validation on new targets. Specifically, we selected five compounds, BaMo\textsubscript{0.3}In\textsubscript{0.7}O\textsubscript{2.95}(BMI), BaNb\textsubscript{0.4}In\textsubscript{0.6}O\textsubscript{2.9}(BNI), Hg{[}B(CN)\textsubscript{4}{]}\textsubscript{2}, CoCo(CN)\textsubscript{6}, and SrNb\textsubscript{2}Fe\textsubscript{2}(PO\textsubscript{4})\textsubscript{6}, that had not been reported in any experimental literature and were designed based on our accumulated experience. These compounds fall into two classes with distinct target properties and structural chemistry. BMI and BNI were designed as proton-conducting electrolytes for intermediate-temperature solid oxide fuel cells. Motivated by Ba-based perovskite oxides, a high concentration of trivalent In\textsuperscript{3+} is introduced on the B site to create a percolated proton-conduction environment, while high-valence Mo and Nb cations are incorporated to suppress the oxygen-vacancy concentration to a moderate level, thereby sustaining framework stability while retaining the vacancies needed for hydration and mobile-proton formation. In contrast, Hg{[}B(CN)\textsubscript{4}{]}\textsubscript{2}, CoCo(CN)\textsubscript{6}, and SrNb\textsubscript{2}Fe\textsubscript{2}(PO\textsubscript{4})\textsubscript{6} were selected as candidate negative-thermal-expansion materials based on the average-atomic-volume concept proposed in our previous work\cite{ref34}, which suggests that compounds with relatively large average atomic volume and open framework structures are more likely to exhibit negative thermal expansion. Accordingly, these three compounds contain polyatomic linking units such as cyanide or phosphate groups, giving them more open and flexible frameworks than dense oxide perovskites. Spanning distinct target applications, chemical compositions, and structural motifs, these five materials provide a stringent test of the generality and effectiveness of PIRAG-LM across markedly different chemical and structural regimes (Fig.~\ref{fig:5}a).

For each target, PIRAG-LM first generated candidate synthesis routes together with precursor sets and processing conditions. Materials experts then evaluated these proposals on the basis of chemical plausibility, experimental accessibility and available equipment, and selected feasible routes for laboratory testing. The resulting synthesis strategies can be grouped into three categories (Fig.~\ref{fig:5}a): direct solid-state synthesis for BaMo\textsubscript{0.3}In\textsubscript{0.7}O\textsubscript{2.95} and BaNb\textsubscript{0.4}In\textsubscript{0.6}O\textsubscript{2.9}; a sol--solid route (solution-phase precursor preparation followed by solid-state calcination) for SrNb\textsubscript{2}Fe\textsubscript{2}(PO\textsubscript{4})\textsubscript{6}; and solution-phase synthesis for CoCo(CN)\textsubscript{6} and Hg{[}B(CN)\textsubscript{4}{]}\textsubscript{2}. The experts further refined the initial conditions during experimental implementation in response to practical considerations and experimental feedback, including precursor reactivity, mixing homogeneity, phase formation and product stability.

Figure~\ref{fig:5}b--f compares the initial PIRAG-LM proposals with the final experimental procedures. Overall, the experimentally implemented routes retained the route class, precursor logic and major processing sequence suggested by the model. The main differences occurred in operational parameters, including calcination temperature, dwell time, milling duration, evaporation temperature and drying temperature. These adjustments reflect experimental optimization rather than a change in the underlying synthesis strategy. The measured PXRD patterns are consistent with the calculated Bragg positions for all five targets, supporting successful formation of the target phases. Thus, PIRAG-LM provides an evidence-grounded starting point that narrows the initial synthesis space, whereas expert assessment and experimental iteration remain necessary to identify workable conditions for each material.

\subsection{Interpretability and evidence analysis}

Interpretability is essential for synthesis planning because predictions must be traceable to evidence that researchers can verify and refine\cite{ref35}. PIRAG-LM achieves this through the material-centered SSKB and RASF: rather than retrieving text passages by semantic similarity, it retrieves previously synthesized materials through chemical, structural, and thermodynamic similarity (CS, SS, and TS). Each retrieved material is linked to route-level records, including precursors, processing conditions, and experimental constraints. Consequently, every recommendation can be traced to explicit physical precedents, explaining both why a precedent was selected and which synthesis knowledge was transferred to the target.

\begin{figure*}[p]
  \centering
  \includegraphics[width=\textwidth,height=0.76\textheight,keepaspectratio]{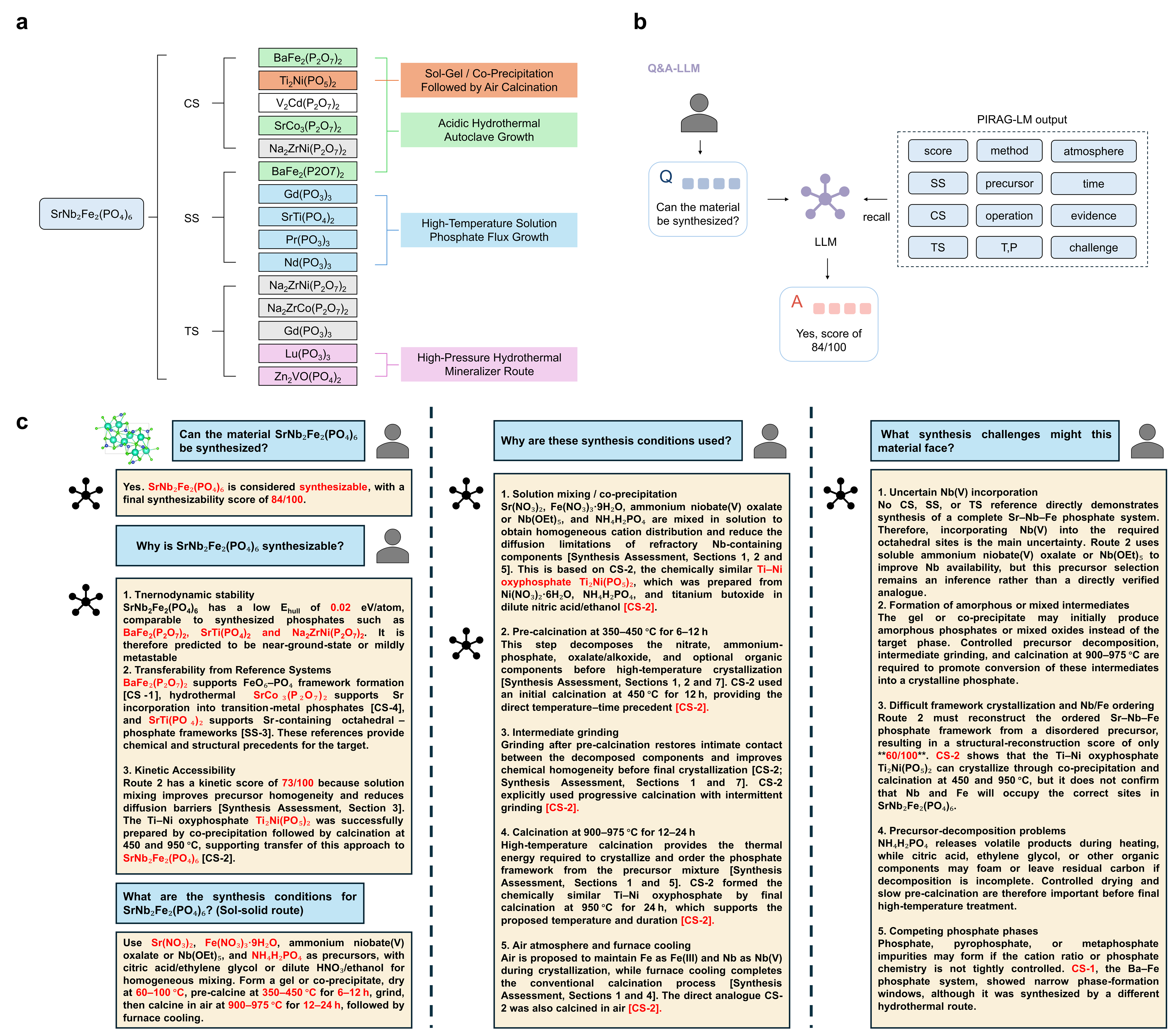}
  \caption{Interpretable, evidence-linked synthesis reasoning in PIRAG-LM. (a) Complementary routes for SrNb\textsubscript{2}Fe\textsubscript{2}(PO\textsubscript{4})\textsubscript{6} generated from chemically, structurally, and thermodynamically similar precedents. (b) Question-answering interface that recalls the synthesizability score, synthesis plan, CS/SS/TS precedents, and predicted challenges. (c) Claim-level evidence chain linking the 84/100 synthesizability assessment, proposed sol--solid conditions, and identified experimental risks to specific retrieved records.}
  \label{fig:6}
\end{figure*}

Figure~\ref{fig:6}a uses synthesis-method selection as an example of SrNb\textsubscript{2}Fe\textsubscript{2}(PO\textsubscript{4})\textsubscript{6} to illustrate how retrieved precedents support PIRAG-LM predictions. RASF retrieves the five closest precedents from each of the CS, SS and TS channels. Ti\textsubscript{2}Ni(PO\textsubscript{5})\textsubscript{2} and related CS precedents support sol--gel/co-precipitation and acidic hydrothermal routes; the phosphate frameworks retrieved through SS support high-temperature solution or phosphate-flux growth; and Lu(PO\textsubscript{3})\textsubscript{3} and Zn\textsubscript{2}VO(PO\textsubscript{4})\textsubscript{2} retrieved through TS support a high-pressure hydrothermal-mineralizer route. Thus, the three physical similarity spaces probe complementary regions of synthesis space, yielding distinct candidate routes. The same evidence-transfer process is used to predict precursors, conditions and potential synthesis challenges.

To make PIRAG-LM reasoning directly interrogable, we developed a Q\&A-LLM that retrieves and organizes its structured outputs---including synthesizability scores, candidate routes, precursors, conditions, physical precedents, and anticipated challenges---into evidence-linked answers rather than post hoc explanations (Fig.~\ref{fig:6}b). For SrNb\textsubscript{2}Fe\textsubscript{2}(PO\textsubscript{4})\textsubscript{6}, the system reports a synthesizability score of 84/100 and recommends a sol--solid route, using sol--gel or co-precipitation for solution-phase precursor preparation followed by staged solid-state calcination. The main anticipated risks are incomplete Nb incorporation, amorphous intermediates, framework reconstruction, and competing phosphate phases (Fig.~\ref{fig:6}c).

When asked why the material is considered synthesizable, the system reconstructs the decision from the recalled physical precedents. First, the target has a low energy above the convex hull of 0.02 eV/atom, indicating that it is close to the thermodynamic ground state. Second, the CS and SS precedents provide complementary evidence for constructing the target framework: BaFe\textsubscript{2}(P\textsubscript{2}O\textsubscript{7})\textsubscript{2}\cite{ref36} supports the formation of Fe--O--P units, whereas SrTi(PO\textsubscript{4})\textsubscript{2}\cite{ref37} and Na\textsubscript{2}ZrNi(P\textsubscript{2}O\textsubscript{7})\textsubscript{2}\cite{ref38} support Sr incorporation and the stabilization of related octahedral phosphate networks. Third, Ti\textsubscript{2}Ni(PO\textsubscript{5})\textsubscript{2} provides a kinetic precedent in which homogeneous precursor mixing followed by staged calcination produces a crystalline phosphate phase\cite{ref39}. Together, these retrieved records support a solution-assisted route involving precursor homogenization, pre-calcination, intermediate grinding and high-temperature crystallization.

The same evidence also defines the limits of this inference. None of the retrieved records directly demonstrates successful Nb(V) incorporation into the complete Sr--Nb--Fe phosphate framework or ordered Nb/Fe site occupation. The positive synthesizability judgement is therefore conditional on knowledge transferred from related systems rather than direct evidence for the target itself. In this process, RASF provides an external evidential scaffold, and the language model reasons over the retrieved material records to connect target properties, transferable synthesis knowledge and candidate routes. Claim-level CS, SS and TS labels make each step inspectable, distinguishing evidence-constrained reasoning from explanations generated solely from the model's parametric memory.

This precedent-to-decision logic is also observed in the experimental validation cases. For BMI, retrieved Ba/Sr--Mo/W and Ba--In--M--O oxides support the use of BaCO\textsubscript{3}, MoO\textsubscript{3} and In\textsubscript{2}O\textsubscript{3}, together with staged heating and prolonged high-temperature treatment. The experiment retained this precursor system and route logic, with the final temperature increased to 1,400~$^\circ$C for kinetic optimization and improved phase purity. Additional detailed evidence chains for BMI, BNI, Hg{[}B(CN)\textsubscript{4}{]}\textsubscript{2} and CoCo(CN)\textsubscript{6} are provided in the Supplementary Information.

\section{Discussion}

In this work, we developed PIRAG-LM, a physics-informed retrieval-augmented language-model framework for inorganic crystal synthesis planning. PIRAG-LM reframes synthesizability from an intrinsic binary property into a route-dependent, evidence-grounded quantity determined by the accessibility of feasible synthesis pathways. It combines a structured synthesis knowledge base containing route-level literature records with chemical, structural, and thermodynamic descriptors to retrieve physically relevant precedents and generate candidate routes, precursor sets, conditions, difficulty assessments, and anticipated challenges. By linking each recommendation to explicit material analogues and their synthesis records, the framework provides interpretable reasoning and enables route-level diagnosis of missing knowledge, including high-pressure or thin-film-specific requirements.

Prospective synthesis of five new compounds---BMI, BNI, Hg{[}B(CN)\textsubscript{4}{]}\textsubscript{2}, CoCo(CN)\textsubscript{6}, and SrNb\textsubscript{2}Fe\textsubscript{2}(PO\textsubscript{4})\textsubscript{6}---demonstrates that PIRAG-LM can connect precursor selection, processing conditions, and structural formation to retrieved evidence, helping translate theoretical crystal predictions into experimentally realized materials. Overall, the framework combines physical descriptors, historical synthesis knowledge, and language-model reasoning to provide a practical foundation for interpretable, route-level synthesis planning in complex inorganic materials.

\clearpage
\section{Methods}

\subsection{Construction of the structured synthesis knowledge base}

We first obtained experimentally synthesized materials from the MP database and collected their associated bibliographic records from the ICSD. Because ICSD records include theoretical studies, characterization-only reports and experimental synthesis studies, retrieved publications were screened to retain only articles reporting explicit experimental syntheses. The retained literature was processed using DeepSeek-V4, which classified synthesis-related information into structured fields, including synthesis method, precursors, temperature, pressure, atmosphere, processing sequence and reported experimental difficulties. The extracted records underwent multiple non-overlapping rounds of review by large language models. Following this automated validation, 500 material structures and their associated synthesis records were manually inspected, and records with overall correct content were incorporated into the dataset.

In addition to experimental synthesis information, the SSKB includes physical descriptors for each material. Crystal structures were converted into natural-language descriptions, and energy above the convex hull, formation energy, electronegativity and band gap were obtained from MP or calculated using \emph{Pymatgen}. These experimental and physical attributes were integrated to form the SSKB used in this work. For the thin-film SSKB, substrate identity was included as an additional route-level field because of its role in determining thin-film phase formation and synthesis conditions.

\subsection{Physics-informed material representations}

Each target material was represented in three complementary physical spaces: chemical similarity (CS), structural similarity (SS) and thermodynamic-state similarity (TS). The inputs were a crystallographic information file (CIF) and an independently calculated energy above the convex hull, $E_{\mathrm{hull}}$, in eV/atom. CIF files were parsed with \emph{Pymatgen}, crystallographic sites closer than 0.1~\AA{} were merged by averaging occupancies and coordinates, and structures that remained disordered or chemically invalid were excluded from retrieval. For CS, the reduced composition was encoded as a 31-dimensional vector. For an element \emph{i} with stoichiometric amount \emph{n\textsubscript{i}}, the atomic fraction \emph{f\textsubscript{i}} was calculated using Eq. (1).

\begin{equation} f_i=\frac{n_i}{\sum_j n_j}. \label{eq:atomic-fraction} \end{equation}

Each element was then described by 14 elemental properties, including atomic number, Pauling electronegativity, atomic mass, periodic-table row and group, atomic radius, average ionic radius, Mendeleev number, oxidation-state statistics and three element-class indicators. The composition-weighted mean and the weighted dispersion of the 11 continuous elemental descriptors were calculated using Eqs. (2) and (3).

\begin{equation} \mathbf{\mu}_p=\sum_i f_i\mathbf{p}_i \in \mathbb{R}^{14}. \label{eq:property-mean} \end{equation}

\begin{equation} \sigma_{p,c}=\left[\sum_i f_i\left(p_{i,c}-\mu_{p,c}\right)^2\right]^{1/2}. \label{eq:property-dispersion} \end{equation}

Together with six global composition statistics---the number of elements, compositional entropy, maximum and minimum atomic fractions, atomic-fraction range and the sum of squared atomic fractions---these quantities define the 31-dimensional CS vector in Eq. (4). Missing elemental properties were imputed using corpus medians, and all features were standardized with the imputer and scaler fitted on the SSKB corpus.

\begin{equation} \mathbf{X}_{\mathrm{CS}}=\left[\mathbf{\mu}_p,\mathbf{\sigma}_p,\mathbf{s}_{\mathrm{comp}}\right]\in\mathbb{R}^{31}. \label{eq:cs-vector} \end{equation}

For SS and TS, structural information was obtained from the latent atomic descriptors of the pretrained MACE-MPA-0 medium foundation interatomic potential, used as a fixed encoder without task-specific fine-tuning. For a structure with \emph{N} atoms, MACE generates a 256-dimensional descriptor \emph{h\textsubscript{k}} for each atom \emph{k}; these descriptors were averaged to obtain the crystal-level vector m in Eq. (5). This representation retains information on elemental identity, local coordination, bonding geometry and the distribution of atomic environments across the unit cell. The same encoder was applied to all 13,820 structures in the searchable index, and the resulting MACE block was standardized feature by feature using parameters fitted on the complete representation corpus.

\begin{equation} \mathbf{m}=N^{-1}\sum_{k=1}^{N}\mathbf{h}_k\in\mathbb{R}^{256}. \label{eq:mace-vector} \end{equation}

Thermodynamic information was encoded separately because a single $E_{\mathrm{hull}}$ value is highly skewed toward zero and does not resolve stable, near-stable and high-energy metastable materials with equal sensitivity. We therefore transformed $E_{\mathrm{hull}}$ into the five-dimensional vector in Eq. (6), which contains the absolute energy, a logarithmic compression of the high-energy tail, an exponential term emphasizing proximity to the convex hull on a 25 meV/atom scale, and two indicator functions for effectively stable (\textless1 meV/atom) and low-energy metastable (\textless50 meV/atom) compounds.

\begin{equation}
\begin{aligned} \mathbf{t}_{E_{\mathrm{hull}}}=\big[&E_{\mathrm{hull}},\,\ln(1+100E_{\mathrm{hull}}),\,\exp(-E_{\mathrm{hull}}/0.025),\\ &\mathbb{I}(E_{\mathrm{hull}}<0.001),\,\mathbb{I}(E_{\mathrm{hull}}<0.05)\big]\in\mathbb{R}^{5}. \end{aligned}
\label{eq:thermodynamic-vector}
\end{equation}

The standardized 256-dimensional MACE vector $\widetilde{\mathbf{m}}$ was used directly as the SS representation, with the thermodynamic block assigned zero weight. For TS, the standardized MACE vector $\widetilde{\mathbf{m}}$ and five-dimensional thermodynamic vector $\widetilde{\mathbf{t}}$ were concatenated with MACE:$E_{\mathrm{hull}}$ block weights of 0.2:1.0. The resulting SS and TS representations are therefore 256 and 261 dimensional, respectively, as defined in Eqs. (7) and (8).

\begin{equation} \mathbf{X}_{\mathrm{SS}}=\widetilde{\mathbf{m}}\in\mathbb{R}^{256}. \label{eq:ss-vector} \end{equation}

\begin{equation} \mathbf{X}_{\mathrm{TS}}=\left[0.2\widetilde{\mathbf{m}},\,1.0\widetilde{\mathbf{t}}\right]\in\mathbb{R}^{261}. \label{eq:ts-vector} \end{equation}

Nearest-neighbour comparisons were performed in the original high-dimensional spaces rather than in UMAP projections. CS candidates were ranked by decreasing cosine similarity in the standardized 31-dimensional chemical space. SS candidates were ranked by increasing cosine distance in the 256-dimensional structural space, whereas TS candidates were ranked in the 261-dimensional fused structural--thermodynamic space; in both cases, the reported similarity was defined as \emph{1 -- d\textsubscript{cos}} according to Eq. (9). Unless otherwise stated, the five nearest neighbours from each space were retained. The two-dimensional UMAP maps were used only for visualization and did not affect retrieval. Finally, the retrieved material identifiers were linked back to route-level SSKB records to generate the \emph{CS\textsubscript{topN}}, \emph{SS\textsubscript{topN}} and \emph{TS\textsubscript{topN}} context files supplied to the language model, including synthesis methods, precursors, processing conditions and source evidence.

\begin{equation} S_{\cos}(\mathbf{u},\mathbf{v})=\frac{\mathbf{u}\cdot\mathbf{v}}{\lVert\mathbf{u}\rVert_2\lVert\mathbf{v}\rVert_2},\qquad d_{\cos}(\mathbf{u},\mathbf{v})=1-S_{\cos}(\mathbf{u},\mathbf{v}). \label{eq:cosine} \end{equation}

\subsection{Evaluation of synthesis predictions}

A method prediction was considered correct when it matched any experimentally reported route family, including solution evaporation, solid-state reaction, cooling crystallization, flux growth, RF co-sputtering, hydrothermal synthesis, mechanochemical ball milling and high-pressure synthesis. Routes representing distinct physical synthesis regimes were not treated as equivalent.

Precursor predictions were evaluated using three criteria. First, precursor coverage measured the fraction of experimentally reported precursors reproduced exactly by the predicted set; chemically equivalent substitutions were not counted as matches:

\begin{equation} C=\frac{N_{\mathrm{matched\ roles}}}{N_{\mathrm{experimental\ roles}}}. \label{eq:coverage} \end{equation}

Second, core-precursor usability was evaluated as a binary material-level criterion. A predicted core set was considered usable when it could provide the required composition and reactive building units under the predicted route. Chemically equivalent alternatives were accepted, whereas sets with missing elemental sources, incompatible reaction pathways or incorrect stoichiometry were rejected.

Chemical substitutability was considered only for core-precursor usability and not for precursor coverage, which required an exact match to the experimentally reported compounds. A predicted core precursor was considered chemically substitutable when it could provide the same target-forming elements or reactive building units with compatible stoichiometry and oxidation states under the same synthesis route. The substitute also had to be compatible with the reported temperature, pressure, atmosphere and reaction medium. Substitutions were rejected if they changed the required elemental balance, introduced incompatible by-products, altered key intermediates, or required a different reaction mechanism or physical synthesis regime.

For example, replacing one soluble metal salt with another was considered acceptable when both generated the same reactive ionic species without changing the solution chemistry. By contrast, Zn\textsubscript{3}N\textsubscript{2} + W was not considered equivalent to the experimentally used Li\textsubscript{6}WN\textsubscript{4} + ZnBr\textsubscript{2} precursors for Zn\textsubscript{3}WN\textsubscript{4} because the latter involves an activated metathesis pathway. Similarly, NH\textsubscript{3} was not considered equivalent to N\textsubscript{2} in the high-pressure synthesis of Sb\textsubscript{3}N\textsubscript{5} because the substitution changes both the nitrogen source and the reaction mechanism. For Cs\textsubscript{3}V\textsubscript{9}Te\textsubscript{13}, Cs\textsubscript{2}Te\textsubscript{3}+V was not treated as equivalent to an elemental Cs/V/Te flux mixture because pre-forming the Cs--Te binary modifies the accessible stoichiometry and reaction environment. These substitutions were therefore classified as core-precursor errors rather than condition-prediction errors.

Third, auxiliary-precursor matching was evaluated only when the experimental route contained identifiable solvents, mineralizers, fluxes, pH regulators, ligands or reactive media. An auxiliary prediction was considered matched when it reproduced the reported species or an equivalent chemical function under the same route. Precursor coverage and core usability were evaluated for all 41 method-correct materials, whereas auxiliary matching was evaluated for the 12 applicable cases.

Reaction-condition prediction was evaluated only after the synthesis method and core precursor set had passed the preceding criteria. Temperature and pressure were compared as numerical intervals rather than by a fixed binary tolerance. Both predicted and experimental values were converted to intervals, with a single reported value treated as an interval whose lower and upper bounds were identical. For each material, the condition error was defined as the shortest distance between the experimental interval and the predicted interval:

\begin{equation}
e=\max\left\{0,\ L_{\mathrm{exp}}-U_{\mathrm{pred}},\ L_{\mathrm{pred}}-U_{\mathrm{exp}}\right\},
\label{eq:interval_error}
\end{equation}

where \(L_{\mathrm{exp}}\) and \(U_{\mathrm{exp}}\) are the lower and upper bounds of the experimental interval, and \(L_{\mathrm{pred}}\) and \(U_{\mathrm{pred}}\) are the corresponding bounds of the predicted interval. Overlapping intervals therefore gave zero error, whereas non-overlapping intervals were penalized by the gap between the nearest interval endpoints. The mean absolute error was calculated as

\begin{equation}
\mathrm{MAE}=\frac{1}{N}\sum_{i=1}^{N} e_i,
\label{eq:condition_mae}
\end{equation}

and was computed separately for temperature and pressure over the eligible materials. When multiple heating or pressure stages were reported, the interval corresponding to the phase-forming step was used. When multiple experimental recipes were available, the recipe consistent with the matched route and core-precursor judgment was used for comparison.

\subsection{Synthesis and characterization}

Five target materials---BMI, BNI, Hg{[}B(CN)\textsubscript{4}{]}\textsubscript{2}, CoCo(CN)\textsubscript{6} and SrNb\textsubscript{2}Fe\textsubscript{2}(PO\textsubscript{4})\textsubscript{6}---were prospectively evaluated before their final experimental procedures were available to PIRAG-LM or included in the SSKB. The target CIFs and available thermodynamic descriptors were supplied as inputs. For each target, PIRAG-LM retrieved five synthesized precedents using chemical, structural and thermodynamic similarities and generated candidate plans specifying the synthesis route, precursors, processing sequence, operating conditions and anticipated difficulties. Experimental researchers selected equipment-compatible plans and optimized the operational parameters according to precursor reactivity, mixing homogeneity, intermediate-phase formation and PXRD feedback.

Phase formation was assessed by powder X-ray diffraction (PXRD) using Cu K$\alpha$ radiation with a Bruker D8 Advance diffractometer (Laguna Hills, CA, USA) by comparing the measured diffraction patterns with Bragg positions calculated from the target structures. A prediction was considered experimentally validated when the implemented procedure retained the predicted route family, core-precursor chemistry and principal processing sequence, and PXRD supported formation of the target phase. Adjustments to milling time, drying or evaporation temperature, calcination temperature and dwell time were treated as condition optimization rather than changes in the predicted synthesis route. The synchrotron XRD (SXRD) data were collected at the BL02B2 beamline (SPring-8, Japan, $\lambda = 0.4500$~\AA{}) and the 11-ID-C beamline (APS, USA, $\lambda = 0.1173$~\AA{}). The crystal structure was refined by the Rietveld method using the FullProf program\cite{ref40}.

\section*{Data availability}

The SSKB dataset developed in this study, including the curated route-level synthesis records and associated chemical, structural, and thermodynamic information, is available at \url{https://github.com/Jweijian/PIRAG-LM}. Data and synthesis-feasibility reports can also be accessed through the PIRAG-LM platform at \url{https://groupcookie.com}.

\section*{Code availability}

The PIRAG-LM code, including the scripts used for data processing, retrieval, synthesis-route generation, and analysis, is openly available at \url{https://github.com/Jweijian/PIRAG-LM}. The online platform is accessible at \url{https://groupcookie.com}.

\section*{Acknowledgements}

We acknowledge funding support by the Innovation Program for Quantum Science and Technology (Grant No. 2024ZD0300102). This work was also supported by the National Natural Science Foundation of China (Grant Nos. 12188101, 12274081, 124B1003, 22379102 and 22471246), the Natural Science Foundation of Henan Province (No. 252300421038) and the Natural Science Foundation of Jiangsu Province (Grant No. BK20251894).

\section*{Author contributions}

The author contributions are defined following the CRediT system.

Wei-Jian Jiang: Conceptualization, Methodology, Software, Validation, Formal analysis, Investigation, Resources, Data curation, Visualization, Writing---original draft, Writing---review \& editing. Ye-Nan Sha: Investigation, Formal analysis, Data curation, Writing---original draft. Hui Guo: Investigation, Data curation. Jie Chen: Formal analysis. Yu-Cai Liang: Formal analysis. Ke Zhou: Writing---review \& editing. Qi-Long Gao: Investigation, Validation, Writing---review \& editing. Dong-Lin Han: Investigation, Validation, Writing---review \& editing. Xin-Gao Gong: Conceptualization, Writing---review \& editing. Wan-Jian Yin: Conceptualization, Methodology, Validation, Resources, Formal analysis, Writing---original draft, Writing---review \& editing.


\end{document}